\documentclass[12pt]{article}
\usepackage{epsfig}
\begin{document}
\title{On the enhancement of the QCD running coupling in the 
 noncontractible space and anomalous TeVatron and HERA data}
\author{D. PALLE \\
Zavod za teorijsku fiziku, Institut Rugjer Bo\v {s}kovi\'{c} \\
P. O. Box 180, 10002 Zagreb, CROATIA}  
\date{ }
\maketitle
 We show that the existence of the fundamental ultraviolet
 cut-off (minimal scale) fixed by  weak interactions enhances the
 QCD running coupling evaluated at one quantum loop level, starting
 at the scale in the vicinity of the cut-off. The enhancement of the
 QCD running coupling
 could completely explain the observed
 anomalous TeVatron and HERA data. 
 The QCD in
 the noncontractible space is not an asymptotically free gauge
 theory. \newline

PACS:  \newline 11.10.Hi Renormalization group evolution and parameters
\newline 12.38.-t Quantum chromodynamics
\newline 12.60.-i Models beyond the standard model
\\

\newpage

\section{Introduction}

We are entering into the era of the very important measurements in
particle physics as well as in cosmology and astrophysics. One
expects the assurance of the results that indicate  the existence
of massive neutrinos and lepton flavour mixing coming from the solar
and atmospheric neutrino data, LSND experiment and from various
astrophysical and cosmological data relevant for measuring  cosmic mass
density and structure formation in the Universe.
The anomalous events in particle physics observed at high energy
hadron-hadron collisions at TeVatron and lepton-hadron collisions
at HERA are especially intriguing.

All these results strongly support the necessity to modify, enlarge or
improve the Standard Model(SM) of particle physics. It has been recently
proposed \cite{r1} a mechanism for the gauge symmetry breaking
without the introduction of the Higgs scalar. The ultraviolet
singularity and the SU(2) global anomaly problems appear as 
milestone points that could lead to the improvement of the SM.
Namely, the embedding of the SU(2) gauge symmetry into the
SU(3) symmetry gives the natural and unique solution 
of the nonperturbative consistency with respect to the SU(2)
anomaly, while the hypothesis of the 
noncontractible space triggers the violation of  gauge, discrete
and conformal symmetries \cite{r1}.

The qualitative analysis of bootstrap equations in the 
nonsingular theory can give the 
insight into the understanding of the problem of a number of fermion families,
 mass gaps between the families, the smallness of neutrino masses, etc.
The lepton number is spontaneously broken and neutrinos
appear as Majorana particles. The neutrino masses are cosmologically
acceptable and confirmed by Super-Kamiokande \cite{r2},
 the heaviest light neutrino could play the role of the hot
dark matter particle \cite{r1,r3} and one of heavy neutrinos could be a
candidate for the cold dark matter \cite{r4}. 
 We are in a position to solve the problem of the baryogenesis through 
 leptogenesis because of the broken lepton number.
 A calculation of the $\eta$-
parameter of the cosmological nucleosynthesis \cite{r3} could cause a
 severe test of the theory.

Introducing into the theory the fundamental scale defined by weak
interactions, as the only fundamental interaction that can provide 
 nonvanishing dimensionfull quantity-the mass of
the weak gauge boson, one has to check the relevance of this scale in the
 gravity and cosmology. We claim \cite{r5} that the weak scale 
 is also a natural fundamental scale in the Einstein-Cartan nonsingular
cosmology where torsion plays a crucial role in preventing
 the appearance of the cosmological singularity. However, the greatest
challenge of the Einstein-Cartan cosmology is the possibility to solve
the problem of the mass density of the Universe and the cosmological
constant problem (without fine-tuning) at the space-like infinity
($T_{\gamma}=0 K$), that means at the time when the Universe is very similar to
 its present evolutionary stage ($T_{\gamma}=2.73 K$) \cite{r5}. 
In addition, the existence of the  
spinning dark matter particles (light and heavy neutrinos)
and the global vorticity of the
Universe are required. \cite{r4,r5,r6}
The EC cosmology can also solve the problem of the primordial
mass density fluctuation \cite{r7}.

It has been also shown that the effect of the fundamental length in quantum
 mechanics \cite{r8} is the spectrum-line broadening that is proportional to the
square of the fundamental length $\Delta E\propto (\frac{d}{R})^{2}$. 
Lee's discrete quantum mechanics (quantum mechanics on the lattice) gives
 different observable phenomena with different bounds and estimates \cite{r9}. 

This paper is devoted to the study of the QCD running coupling
in the noncontractible space at one quantum loop and its comparison with
the SM calculations. In the next section we present the perturbative
calculation  supplied with all the necessary details
in the Appendix.  In the concluding section we outline numerical
results and discuss their relevance with respect to the recently
observed anomalous events at TeVatron and HERA.

\section{Perturbative calculus of the QCD running coupling}

The UV cut-off is fixed in a gauge and Lorentz invariant
 manner applying the Wick's theorem in the trace anomaly \cite{r1}.  Contrary to
 other scale fixing procedures, such as in the nonlocal gauge theory 
through the nonuniversal functionals, the relation for the weak boson
mass is similar to that of the Higgs mechanism but now instead of the 
vacuum expectation value of the scalar field figures the universal
cut-off (modulo real number),
 thus defined by the gauge and Lorentz invariant quantities, namely
 the weak boson mass and the weak coupling constant
 \cite{r1}: $\Lambda=
\frac{2\pi}{\sqrt{6}g}M_{W}=\frac{\hbar}{c d }$.
 
We can use all formalisms of the local relativistic
quantum gauge field theory for the broken (QFD) and the
unbroken (QCD) phase of the theory. The above relation should be
preserved to all orders in perturbation theory and it should be
considered as a definition of the universal fundamental scale .
Operator gauge- and Heisenberg-algebras are intact by this consideration,
 no new operators emerge and one can use all the benefits of the BRST symmetry,
 such as the generalized Ward-Takahashi and Slavnov-Taylor identities for
 the Green's functions and
 the renormalizability of the non-Abelian gauge theory
 \cite{r1}.

 The calculations will be performed in the 't Hooft-Feynman gauge
 with constant nonvanishing quark masses. 
We choose the definition of the running coupling originating from the
  light quark-gluon 
 vertex \cite{r10}.

The momentum subtraction renormalization scheme \cite{r11}  appears as the
naturally suitable scheme for the UV finite theory and we shall apply it
 to the QCD, with and without the fundamental scale.
 
  The following conventions are
adopted for the renormalization constants \cite{r12} :

\begin{eqnarray}
\alpha = Z^{-1}_{\alpha} \alpha _{0},\ Z_{\alpha} = 
(\frac{Z_{1F}}{Z_{2F}})^{2} Z^{-1}_{3YM},\hspace*{23 mm} \nonumber\\
G^{\mu}_{a0}(x) = Z^{1/2}_{3YM} G^{\mu}_{a}(x),\ 
q^{A}_{\alpha 0}(x) = Z^{1/2}_{2F} q^{A}_{\alpha}(x),\hspace*{10 mm} \\
g_{0F} = Z_{1F} Z^{-1/2}_{3YM} Z^{-1}_{2F} g_{F},\ 
\beta(\alpha) = Z_{\alpha}\mu \frac{d Z^{-1}_{\alpha}}{d \mu} = 
\alpha^{-1} \mu \frac{d \alpha}{d \mu}.\nonumber
\end{eqnarray}

The off-mass-shell renormalization conditions define the following  
 physical (renormalized) Green's functions:

\begin{eqnarray}
S^{R}(\not p)_{p^2=-\mu^{2}} = S_{F}(\not p)_{p^2=-\mu^{2}},\hspace*{16 mm}
 \nonumber \\
\ \Gamma^{R}_{\nu}(p,q)_{p^{2}=q^{2}=-\mu^{2}}=
\gamma_{\nu} , \hspace*{25 mm} \\
 asymmetric\ condition:\ p^{2}=q^{2}\neq (p+q)^{2}. \hspace*{10 mm}
 \nonumber
\end{eqnarray}
 
To insure the SU(3) gauge invariance we impose the on-mass-shell
renormalization condition for the polarization operator of the gluon field
 \cite{r13}:

\begin{eqnarray}
\Pi^{on}_{R}(p)_{p^{2}=0}=0 .
\end{eqnarray}

The above conditions define the infinite and finite parts of the
renormalization constants in the SM and the finite renormalization 
constants in the UV-finite theory. 

We have now to relate renormalization constants of the polarization
operator in two distinct (off- and on-mass-shell) renormalization
schemes:

\begin{eqnarray}
\Pi^{on}_{R}(p,m_{i},\Lambda)=Z^{on}_{3YM}(p,m_{i},\Lambda)
 \Pi_{0}(p,m_{i},,\Lambda),\hspace{30 mm} \nonumber \\
\Pi^{off}_{R}(p,m_{i},\mu,\Lambda)=Z^{off}_{3YM}(\mu,m_{i}, 
\Lambda) \Pi_{0}(p,m_{i},\Lambda), \hspace{30 mm} \\
\ \Lambda=fundamental\ UV\ cut-off,\mu= scale\ 
parameter.  \hspace*{25 mm} \nonumber
\end{eqnarray}

The evaluation of the $\beta$-function requires the knowledge of the derivative of the renormalization constant with respect to the scale variable:
 
\begin{eqnarray}
 \frac{\partial Z^{off}_{3YM}(\mu,m_{i},\Lambda)}{\partial ln \mu}=
 \frac{\partial
 \Pi_{R}^{off}(p,m_{i},\mu,\Lambda)}{\partial ln \mu}.
\end{eqnarray}

Because of the universality of the $\beta$-function to the one-loop
 order and Eq.(4), the following relation must be fulfilled:

\begin{eqnarray}
\frac{\partial \Pi^{off}_{R}(p,m_{i},\mu,\Lambda)}
{\partial ln \mu}=\frac{\partial \Pi^{on}_{R}(\mu^{2}=-\frac{1}{p^{2}},
m_{i},\Lambda)}{\partial ln \mu} .
\end{eqnarray}
 
By the choice for the scale variable $\mu^{2}=-\frac{1}{p^{2}}$ in the
on-mass-shell scheme, it is possible to compare the physical
quantities at various spacelike points up to the spacelike infinity.
It is in accordance with the on-mass-shell renormalization condition
 at $p^{2}=0$ for the polarization operator. Thus, we can conclude that:

\begin{eqnarray}
\frac{\partial Z^{off}_{3YM}(\mu,m_{i},\Lambda)}{\partial ln \mu}=
-\frac{\partial \Pi^{on}_{R}(\mu^{2}=-p^{2},m_{i},\Lambda)}
{\partial ln \mu} .  
\end{eqnarray}
  
One can immediately evaluate (see Ref.\cite{r12} or any textbook on the QCD)
 the necessary renormalization constants 
from the quark-gluon vertex, quark and 
 gluon self-energy diagrams in the $\ $'t Hooft-Feynman gauge in terms of
 one-, two- and three-point Green's functions(see Appendix):

\begin{eqnarray}
Z_{i} = 1 + \delta Z_{i}, \hspace*{55 mm} \nonumber \\
\delta Z_{1F}=\frac{\alpha_{s}}{4 \pi}[\frac{1}{6}(2 B_{0}(4p^{2};0,m_
{q})-4 C_{2}^{a}(p,2 p;m_{q},m_{q})+2 p^2 C_{0}(p,2 p;m_{q},m_{q})\nonumber\\
+2 p^{2} C_{1}(p,2 p;m_{q},m_{q}))-\frac{3}{2}(2 B_{0}(4p^{2};0,m_{q})+
p^{2} C_{1}(p,-p;0,m_{q})\nonumber \\
+4 C_{2}^{a}(p,-p;0,m_{q})-p^{2} C_{0}(p,-p;0,m_{q}))]_{p^{2}=-\mu^{2}},
  \hspace*{21 mm} \nonumber \\
\delta Z_{2F}=-\frac{\alpha_{s}}{4 \pi}\frac{8}{3}[B_{0}(-\mu^{2};
0,m_{q})+B_{1}(-\mu^{2};0,m_{q})], \hspace*{20 mm} \nonumber  \\
\delta Z_{3YM}=\frac{\alpha_{s}}{4 \pi}[-\frac{1}{3}\sum_{f}
(2 B_{0}(-\mu^{2};m_{f},m_{f})-\frac{4 m_{f}^{2}}{\mu^{2}}
(B_{0}(-\mu^{2};m_{f},m_{f}) \nonumber\\
-B_{0}(0;m_{f},m_{f}))+5 B_{0}(-\mu^{2};0,0)]+\chi (m_{i},
 \Lambda).
\end{eqnarray}

From the standard definition of the $\beta$ function \cite{r12} we can easily
 find the relation for the QCD running coupling to one quantum loop:

\begin{eqnarray}
\beta(\alpha_{s})\simeq \alpha_{s} \beta_{1}(\mu);\ 
-\frac{1}{\alpha_{s}(\mu)}+\frac{1}{\alpha_{s}(\mu_{0})}=
\int^{\mu}_{\mu_{0}}\frac{\beta_{1}(\kappa)}{\kappa}d \kappa, \hspace*{10 mm}
 \nonumber \\
\beta_{1}(\mu) \equiv \mu \frac{d \Phi(\mu)}{d \mu}, \hspace*{30 mm}
\nonumber \\
\alpha_{s}(\mu)=\frac{\alpha_{s}(\mu_{0})}
{1+\alpha_{s}(\mu_{0})(\Phi(\mu_{0})-\Phi(\mu))},\hspace*{20 mm} \\
\Phi(\mu)=\alpha_{s}^{-1}(-2 \delta Z_{1F}+2 \delta Z_{2F}+
\delta Z_{3YM}).
 \hspace*{13 mm} \nonumber
\end{eqnarray}

Eqs. (8) and (9) give immediately  the standard relation for
the QCD running coupling in the SM with massless quarks:

\begin{eqnarray}
\Phi(\mu_{0})-\Phi(\mu)=\frac{11-\frac{2}{3}n_{f}}{2 \pi} 
ln \frac{\mu}{\mu_{0}}. \nonumber
\end{eqnarray} 
 
To derive the above formula we used the following relations:
\begin{eqnarray}
\frac{\partial B^{\infty}_{0}(-\mu^{2};0,0)}{\partial ln \mu}=-2,\ 
\frac{\partial[ \mu^{2} C^{\infty}_{0}(-\mu^{2};0,0)]}{\partial
 ln \mu}=2.
\end{eqnarray}

Throughout the paper the superscripts "$\infty$" or "$\Lambda$" denote
 the physical quantities evaluated in the standard way or with the
 covariant UV-cut-off $\Lambda$.

Before turning to the numerical study of our basic result Eq.(9),
we should comment three important points: (1) to preserve the gauge
invariance in the case of $\Lambda < \infty$, it is essential to
fulfil condition of Eq.(3) by which ${\cal O}(\Lambda ^{2})$
terms are subtracted away, (2) the dependence of the observables on
the covariant spacelike cut-off $\Lambda$ is completely hidden in
the integration region of the scalar integrals; one should not confuse
this cut-off with some regularization cut-off because
for the theory with $\Lambda < \infty$ there is a unique integration
and a nontrivial analytical continuation procedure for timelike external
momenta (for details see Appendix), (3) the scaling variable $\mu$
can acquire arbitrary value (it is not limited by the cut-off) because
even for $\Lambda < \infty$ the theory is a local gauge theory.

\section{Results and discussion}

We can now illustrate the effect of the fundamental UV cut-off on the
 QCD running coupling, applying Eqs. (8) and (9)
 to the Green's functions with and without the UV cut-off. To make 
a comparison we choose the following set 
of the initial conditions and quark massess\cite{r14,r15} ($\alpha_{s}^{\infty}
\equiv \alpha_{s}(SM)$) :

\begin{eqnarray*}
 Input\ parameters\ for\ Eq.(9): \hspace*{66 mm} \\
\Lambda=326\ GeV,\ \alpha_{s}(\mu_{0})=0.12,\ \mu_{0}=M_{Z}=91.19\ GeV,\ n_{f}=6, 
 \hspace*{20 mm} \\
m_{u}=6\ MeV,\ m_{d}=9\ MeV,\ m_{s}=160\ MeV, \hspace*{40 mm}\\
m_{c}=1.5\ GeV,\ m_{b}=4.5\ GeV,\ m_{t}=175\ GeV \hspace*{37 mm} \\
\end{eqnarray*}

In Figure 1 one can notice the enhancement of the running coupling
  $\alpha^{\Lambda}_{s}$ in comparison with $\alpha^{\infty}_{s}$,
 starting at the scale in the vicinity of the UV cut-off. We have  
  displayed $\alpha_{s}^{2}$ values because the differential cross
 sections
 of various hadron-hadron collisions are proportional
 to $\alpha_{s}^{2}$.
 The enhancement of the inclusive jet cross section at high $E_{T}$ and 
 the excess in the production of W(Z) plus one jet  
 are observed 
 at TeVatron \cite{r16,r17}. 

\begin{figure}
\epsfig{figure=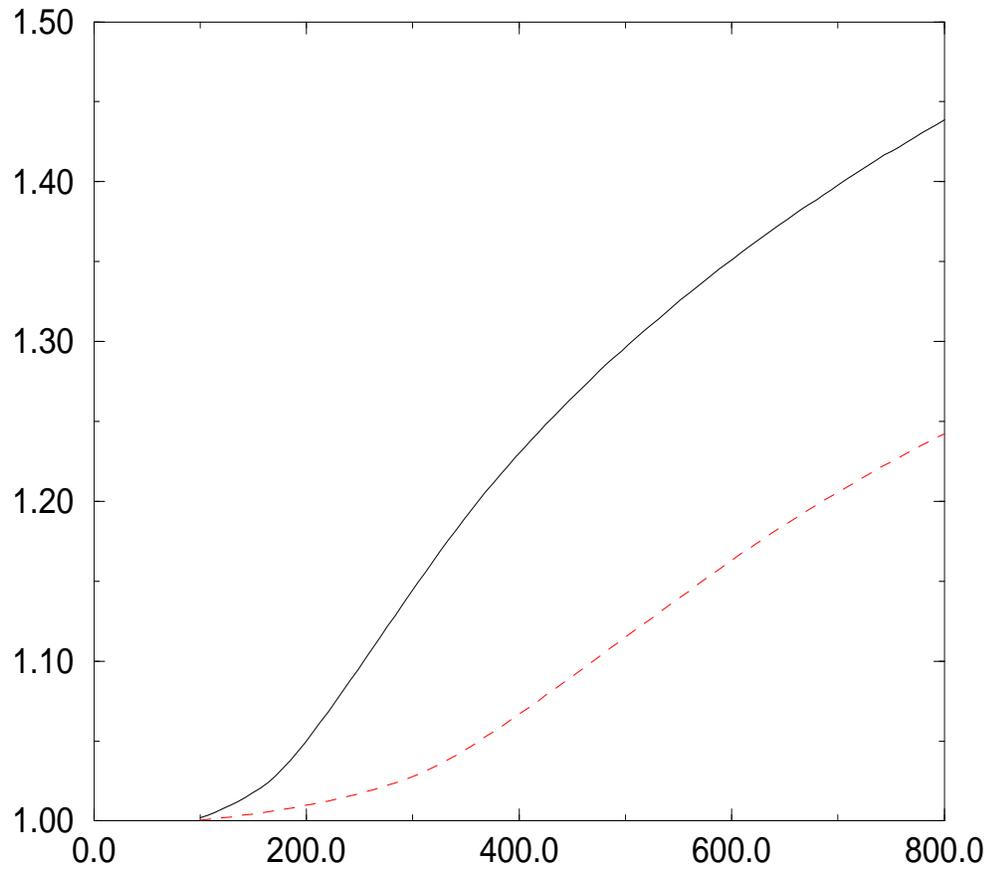,height=140 mm,width=150 mm}
\caption{Solid [dashed] line denotes $(\alpha^{\Lambda}_{s}
/\alpha^{\infty}_{s})^{2}$ vs. $\mu (GeV)$ for $\Lambda$=
326 [600] GeV.}
\end{figure}

In order to show the sensitivity of the results on the magnitude of
the fundamental UV cut-off, one can observe in Figure 1 the smaller effect for
larger cut-off $\Lambda$.

The effects of running masses or two loop corrections
cannot alter our conclusion on the persistent enhancement
of $\alpha_{s}^{\Lambda}(\mu)$ for $\mu \geq 200 GeV$.

The leading order calculation of the Altarelli-Parisi equations
\cite{r18,r19} shows that there is a very small enhancement of parton
distribution functions for small x and very small suppression
for large x at $Q \geq 200 GeV$ (see Figs. 2 and 3).

\begin{figure}
\epsfig{figure=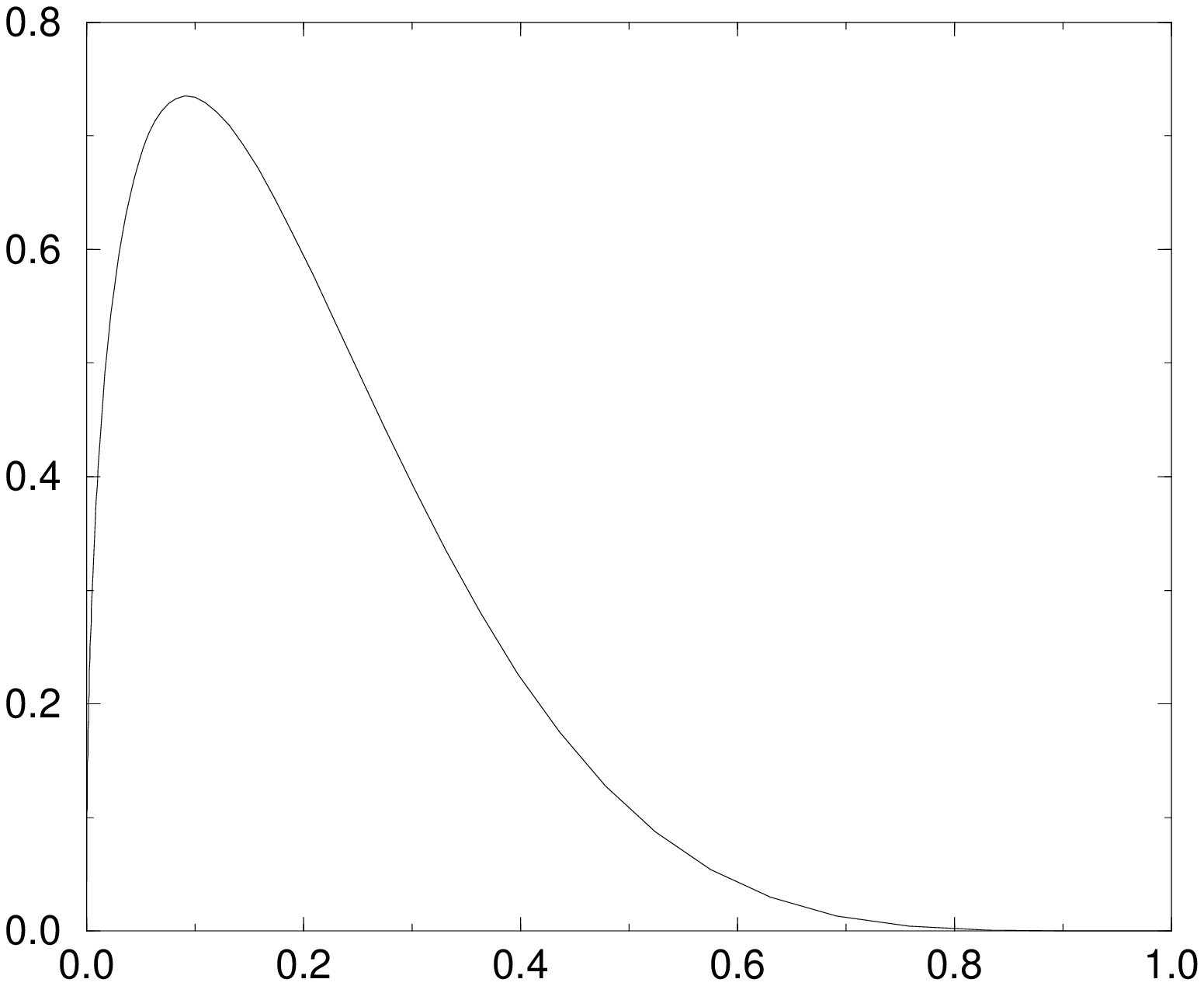,height=140 mm,width=150 mm}
\caption{Solid line denotes $xF_{1}^{\Lambda}(x,Q^{2})$ vs. x for
Q=300 GeV and $\Lambda$=326 GeV.}
\end{figure}

\begin{figure}
\epsfig{figure=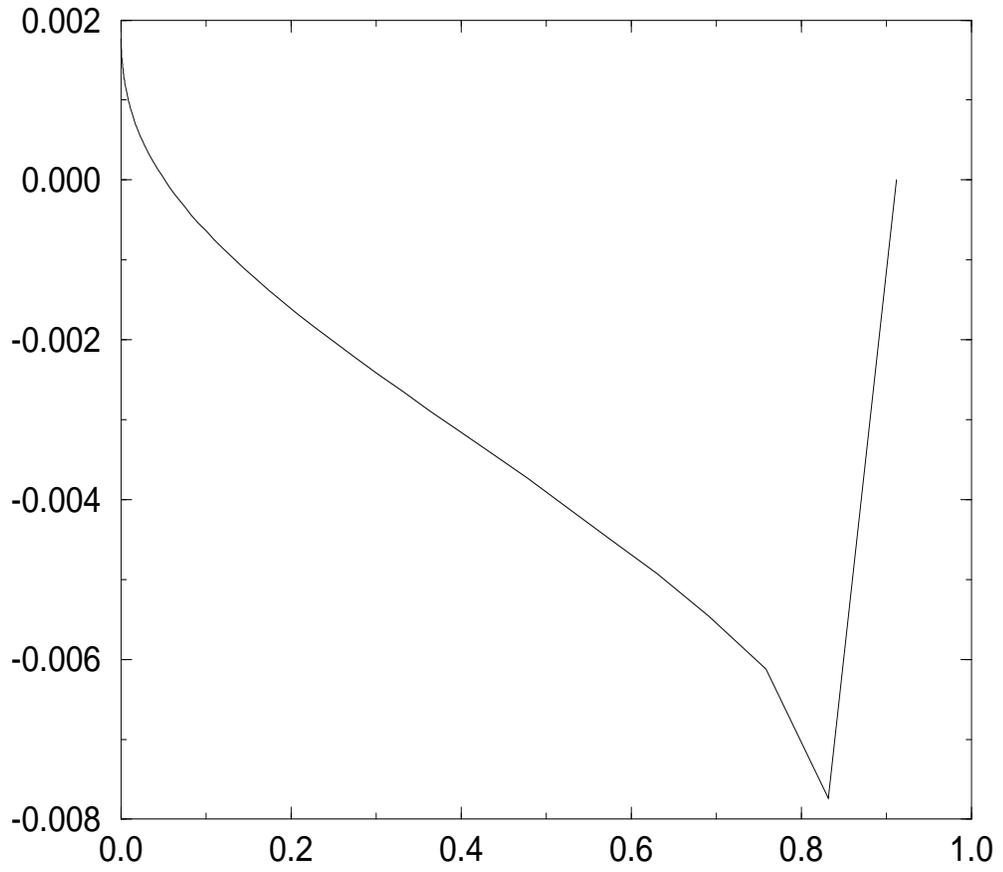,height=140 mm,width=150 mm}
\caption{Solid line denotes $(xF_{1}^{\Lambda}-
xF_{1}^{\infty})/(xF_{1}^{\infty})$ vs. x for
Q=300 GeV.}
\end{figure}

QCD corrections to the electroweak couplings could generate
enhancement above the scale $\mu \simeq 200 GeV$. This
effect is observed at HERA and LEP 2 \cite{r20,r21}:

\begin{eqnarray*} 
|V_{EW+QCD}(q^{2})|^{2} = |V_{EW}(q^{2})|^{2} 
[1+\frac{\alpha_{s}(q^{2})}{\pi} + 
{\cal O}(\alpha^{2}_{s}(q^{2}))] ,
\end{eqnarray*}
\begin{eqnarray*}
enhancement\ factor =\ [1+\frac{\alpha_{s}^{\Lambda}(q^{2})}{\pi}]/
[1+\frac{\alpha_{s}^{\infty}(q^{2})}{\pi}] .
\end{eqnarray*} 

To conclude, one can say that the effect of the noncontractible space
 in QCD is the nonresonant and universal enhancement of various cross 
 sections in 
  $p \bar{p}$ and $ep$ collisions (this conclusion is verified in the
 region where one can apply the perturbative calculus),
 starting at the scale in the vicinity 
 of the UV cut-off. The characteristics of the anomalous TeVatron  
  and HERA data are in accordance with this claim \cite{r16,r20}. 
 Above the scale of $\mu \simeq 500\ GeV$ the QCD coupling $\alpha^{\Lambda}
 _{s}$ is frozen at some nonvanishing value, for example 
 $\lim_{\mu \rightarrow \infty} \alpha^{\Lambda}_{s}(\mu)=0.11$ with parameters
 of Figure 1. The enhancement at the scale relevant at LHC is:
 $(\frac{\alpha^{\Lambda}_{s}}{\alpha^{\infty}_{s}})^{2}(\mu = few\ TeV) \simeq
 2-4$.
 Evidently, the QCD in the noncontractible space is not an 
 asymptotically free gauge field theory \cite{r22}.

\section*{Acknowledgements}
This work was supported by the Ministry of Science and Technology of
the Republic of Croatia under Contract No. 00980103.
It is a pleasure also to thank Prof. O. Nachtmann for his kind hospitality 
and useful discussions during my stay in Heidelberg, 
and to the Alexander von Humboldt-Stiftung 
 for the partial financial support.
\newline

\section*{Appendix}

We use the following definitions and settings of the Green's 
functions with the UV cut-off ($\Lambda$ superscript) and the SM ones($\infty$ superscript)

\begin{eqnarray*}
A(m)=-\frac{1}{\imath\pi^{2}}\int d^{4}q
\frac{1}{q^{2}-m^{2}+\imath\varepsilon},  \\
\end{eqnarray*}
\begin{eqnarray*}
B_{0}(p^{2};m_{1},m_{2})=\frac{1}{\imath\pi^{2}}\int d^{4}q
\frac{1}{(q^{2}-m_{1}^{2}+\imath\varepsilon)
((q+p)^{2}-m_{2}^{2}+\imath\varepsilon)},  \\
\end{eqnarray*}
\begin{eqnarray*}
p_{\mu}B_{1}(p^{2};m_{1},m_{2})=\frac{1}{\imath\pi^{2}}
\int d^{4}q\frac{q_{\mu}}{(q^{2}-m_{1}^{2}+
\imath\varepsilon)((q+p)^{2}-m_{2}^{2}+\imath\varepsilon)},  \\
\end{eqnarray*}
\begin{eqnarray*}
C_{0}(p_{1},p_{2};m_{1},m_{2})=\frac{1}{\imath\pi^{2}}\int d^{4}q
\frac{1}{(q^{2}+\imath\varepsilon)((q+p_{1})^{2}-m_{1}^{2}+
\imath\varepsilon)((q+p_{2})^{2}-m_{2}^{2}+\imath\varepsilon)}, \\
\end{eqnarray*}
\begin{eqnarray*}
p_{1\mu}C_{1}(p_{1},p_{2};m_{1},m_{2})=\frac{1}{\imath\pi^{2}}\int
d^{4}q\frac{q_{\mu}}{(q^{2}+\imath\varepsilon)((q+p_{1})^{2}-
m_{1}^{2}+\imath\varepsilon)((q+p_{2})^{2}-m_{2}^{2}+
\imath\varepsilon)}, \\
\end{eqnarray*}
\begin{eqnarray*}
g_{\mu\nu}C^{a}_{2}(p_{1},p_{2};m_{1},m_{2})+
p_{1\mu}p_{1\nu}C^{b}_{2}(p_{1},p_{2};m_{1},m_{2})= \\
\end{eqnarray*}
\begin{eqnarray*}
=\frac{1}{\imath\pi^{2}}\int d^{4}q \frac{q_{\mu}q_{\nu}}
{(q^{2}+\imath\varepsilon)((q+p_{1})^{2}-m_{1}^{2}+\imath\varepsilon)
((q+p_{2})^{2}-m_{2}^{2}+\imath\varepsilon)}, 
\end{eqnarray*}
\begin{eqnarray*}
2 p^{2} B_{1}(p^{2};m_{1},m_{2})=A(m_{2})-A(m_{1})+
(m_{2}^{2}-m_{1}^{2}-p^{2})B_{0}(p^{2};m_{1},m_{2}), \\
2p_{1}^{2}C_{1}(p_{1},p_{2};m_{1},m_{2})=
B_{0}(p_{2}^2;0,m_{2})-B_{0}((p_{2}-p_{1})^{2};m_{1},m_{2}) \\
+(m_{1}^{2}-p_{1}^{2})C_{0}(p_{1},p_{2};m_{1},m_{2}),\hspace*{16 mm}\\
\end{eqnarray*}
\begin{eqnarray*}
C_{2}^{a}=\frac{1}{3}(\Delta_{1}-\Delta_{2}),\ 
C_{2}^{b}=\frac{1}{3p_{1}^{2}}(4 \Delta_{2}-\Delta_{1}),\hspace*{42 mm} \\
\Delta_{1}=B_{0}((p_{2}-p_{1})^{2};m_{1},m_{2}),\hspace*{42 mm}  \\
\Delta_{2}=\frac{1}{2}[B_{0}((p_{2}-p_{1})^{2};m_{1},m_{2})+
\kappa B_{1}(p_{2}^{2};0,m_{2})+(1-\kappa)B_{1}((p_{2}-p_{1})^{2};
m_{1},m_{2}) \\
+(m_{1}^{2}-p_{1}^{2})C_{1}(p_{1},p_{2};m_{1},m_{2})],\hspace*{77 mm} \\
  p_{2\mu}=\kappa p_{1\mu}, \hspace*{60 mm} 
\end{eqnarray*}
\begin{eqnarray*}
A^{\Lambda}(m)=\Lambda^{2}-m^{2} ln(\frac{\Lambda^{2}+m^{2}}{m^{2}}), \hspace{30 mm} 
\end{eqnarray*}
\begin{eqnarray*}
B_{0}^{\Lambda}(p^{2};m_{1},m_{2}) = \frac{1}{2}[\tilde{B}_{0}^{\Lambda}
(p^{2};m_{1},m_{2})+\tilde{B}_{0}^{\Lambda}(p^{2};m_{2},m_{1})],
\end{eqnarray*}
\begin{eqnarray*}
Re \tilde{B}_{0}^{\Lambda}(p^{2};m_{1},m_{2})=
(\int_{0}^{\Lambda^{2}}d y K(p^{2},y)+\theta (p^{2}-m_{2}^{2}) 
 \int_{-(\sqrt{p^{2}}-m_{2})^{2}}
^{0}d y \Delta K(p^{2},y) )\frac{1}{y+m_{1}^{2}}, \\
K(p^{2},y)=\frac{2 y}{-p^{2}+y+m_{2}^{2}+
\sqrt{(-p^{2}+y+m_{2}^{2})^{2}+4 p^{2} y}},  \hspace*{40 mm}\\
\Delta K(p^{2},y)=\frac{\sqrt{(-p^{2}+y+m_{2}^{2})^{2}+4 p^{2} y}}{p^{2}}.
 \hspace*{60 mm}
\end{eqnarray*}

The integration in the second term \cite{r23} is performed from the branch 
point of the square root $\sqrt{(-p^{2}+y+m_{2}^{2})^{2}+4 p^{2} y}\equiv 
\imath Z$ and the additional kernel is derived as the difference:
$ \Delta K(p^{2},y)=K(p^{2},y)-K^{*}(p^{2},y)=\frac{2 y}{-p^{2}+y+
m_{2}^{2}+\imath Z}-\frac{2 y}{-p^{2}+y+m_{2}^{2}-\imath Z}$.

The integration over singularities is supposed to be the principal value 
integration.

\begin{eqnarray*}
C_{0}^{\Lambda}(p_{1}^{2},(p_{2}-p_{1})^{2},p_{2}^{2};
0,m_{1}^{2},m_{2}^{2}) = 
 \frac{1}{3}[\tilde{C}_{0}^{\Lambda}(p_{1}^{2},
(p_{2}-p_{1})^{2},p_{2}^{2};0,m_{1}^{2},m_{2}^{2}) \\
+\tilde{C}_{0}^{\Lambda}((p_{2}-p_{1})^{2},p_{2}^{2},p_{1}^{2};
m_{1}^{2},m_{2}^{2},0)+\tilde{C}_{0}^{\Lambda}(p_{2}^{2},
p_{1}^{2},(p_{2}-p_{1})^{2};m_{2}^{2},0,m_{1}^{2})],
\end{eqnarray*}
\begin{eqnarray*}
Re \tilde{C}_{0}^{[\Lambda;\infty]}(123)=-\frac{4}{\pi}
\int^{[\Lambda;\infty ]}_{0} dq q  \hspace*{20 mm}\\
\times \int^{+1}_{-1} dx \sqrt{1-x^{2}}
\frac{1}{q^{2}+p_{1}^{2}+m_{1}^{2}+2 q p_{1} x}
\frac{1}{q^{2}+p_{2}^{2}+m_{2}^{2}+2 q p_{2} x},\\
p_{1}^{\mu}=(0,0,0,p_{1}),\ p_{2}^{\mu}=(0,0,0,p_{2}), \hspace*{40 mm}
\end{eqnarray*}
\begin{eqnarray*}
similarly\ for\ \tilde{C}_{0}^{\Lambda}(231)\ and\ \tilde{C}_{0}
^{\Lambda}(312).
\end{eqnarray*}

Symmetrization over external momenta is included in order to
restore the momentum-exchange symmetry when $\Lambda < \infty$
(broken scale symmetry).

The integrals for high momenta up to infinity should be performed
 after the inverse mapping of the integration variable.
For massive quarks and off-shell external momenta Green's functions
 are infrared convergent \cite{r24}. 

In the case of the two-point Green's function $B_{0}^{\Lambda}$
 we need the explicit form of the additional term for the integration
in the timelike region because the integration 
in the spacelike region in the limes $\Lambda 
 \rightarrow \infty$ is divergent. However, the three-point scalar Green's
functions are UV-convergent and we do not need to know the explicit
form of the additional terms because they do not depend on the 
UV cut-off and we can use the analytical continuation of the standard
 Green's functions written in terms of the dilogarithms\cite{r25}:

\begin{eqnarray*}
Re C_{0}^{\Lambda}(p_{i},m_{j})=\int_{0}^{\Lambda^{2}}dq^{2} \Phi 
(q^{2},p_{i},m_{j})
+\int_{TD}dq^{2} \Xi (q^{2},p_{i},m_{j}), \hspace*{10 mm}\\
Re C_{0}^{\Lambda}(p_{i},m_{j})=Re C_{0}^{\infty}(p_{i},m_{j})-
\int^{\infty}_{\Lambda^{2}}d q^{2} \Phi (q^{2},p_{i},m_{j}), 
 \hspace*{20 mm} \\
\Phi \equiv function\ derived\ by\ the\ angular\ integration\ 
after\ Wick's\ rotation,\\
C_{0}^{\infty}\equiv standard\ 't\ Hooft-Veltman\ scalar\ function, 
\hspace*{20 mm} \\
TD\equiv timelike\ domain\ of\ integration. \hspace*{40 mm}
\end{eqnarray*}

\end{document}